\documentclass[aps,prl,twocolumn,superscriptaddress]{revtex4}
\usepackage{graphicx}
\usepackage{subfigure}
\usepackage{amsmath}
\usepackage{amssymb}
\usepackage{bm}
\usepackage{mathrsfs}
\usepackage{hyperref}
\hypersetup{colorlinks=true, linkcolor=black, citecolor=black, urlcolor=black}
\usepackage[english]{babel}
\usepackage{subfigure}
\usepackage{braket}

\newcommand{\m}{\vspace{0.2in}}
\newcommand{\mm}{\vspace{0.09in}}

\begin{document}

\title{Detecting topological transitions in two dimensions by Hamiltonian 
evolution}
\author{Wei-Wei Zhang}
\affiliation{State Key Laboratory of Networking and Switching Technology, Beijing University of Posts and Telecommunications, Beijing 100876, China}
\affiliation{Hefei National Laboratory for Physical Sciences at Microscale, University of Science and Technology of China, Hefei, Anhui 230026, China}
\affiliation{Institute for Quantum Science and Technology, and Department of Physics and Astronomy, University of Calgary, Calgary, Alberta, Canada T2N 1N4}
\affiliation{Centre for Engineered Quantum Systems, School of Physics, The University of Sydney, Sydney, Australia}
\author{Barry C.~Sanders}
\affiliation{Hefei National Laboratory for Physical Sciences at Microscale, University of Science and Technology of China, Hefei, Anhui 230026, China}
\affiliation{Institute for Quantum Science and Technology, and Department of 
Physics and Astronomy, University of Calgary, Calgary, Alberta, Canada T2N 1N4}
\affiliation{Shanghai Branch, CAS Center for Excellence and Synergetic Innovation Center in Quantum Information and Quantum Physics, University of Science and Technology of China, Shanghai 201315, China}
\affiliation{Program in Quantum Information Science, Canadian Institute for Advanced Research, Toronto, Ontario, Canada M5G 1M1}
\author{Simon Apers}
\affiliation{SYSTeMS, Ghent University IR08, Technologiepark 
913, B-9052 Zwijnaarde, Belgium}
\author{Sandeep K.~Goyal}
\affiliation{Indian Institute of Science Education and Research, Mohali, Punjab, 140306 India}
\author{David L. Feder}
\email[Corresponding author: ]{dfeder@ucalgary.ca}
\affiliation{Institute for Quantum Science and Technology, and Department of 
Physics and Astronomy, University of Calgary, Calgary, Alberta, Canada T2N 1N4}

\begin{abstract}
We show that the evolution of two-component particles governed by a 
two-dimensional spin-orbit lattice Hamiltonian can reveal transitions between 
topological phases. A kink in the mean width of the particle distribution 
signals the closing of the band gap, a prerequisite for a quantum phase 
transition between topological phases. Furthermore, for realistic and
experimentally motivated Hamiltonians the density profile in topologically 
non-trivial phases displays characteristic rings in the vicinity of the origin 
that are absent in trivial phases. The results are expected to have immediate 
application to systems of ultracold atoms and photonic lattices.
\end{abstract}
\maketitle


Topological phases have many unusual and potentially useful electronic 
properties, and have been proposed for fault-tolerant quantum computation 
and quantum memories~\cite{Moore2010,Hasan2010,Ryu2010,Qi2011,Pachos2014}.
In one dimensional systems, all topological states can be 
classified~\cite{Chen2011}. In higher dimensions, non-interacting systems can 
be classified in terms of topological invariants such as Chern 
numbers~\cite{Chiu2016}, and much work has been expended in recent years 
attempting to extend this classification to interacting 
systems~\cite{Chen2012,Wang2014,Wang2015a,Wang2015b}. The experimental 
determination of topological invariants in bulk condensed matter systems with 
time-reversal symmetry is not straightforward, however; topological order would 
generally be inferred from the existence of edge 
states~\cite{Hasan2010,Wu2016}. In this work, the presence of non-trivial 
topological order is inferred from particle dynamics.

The exceptional control of integrated photonic and ultracold atomic systems 
makes them ideal testbeds for the production and detection of topological 
order~\cite{Haldane2008,Lu2014,Goldman2016}. After the first realization of the 
photonic analog of the quantum Hall effect~\cite{Wang2009}, topological edge 
modes were observed in both static and driven photonic 
lattices~\cite{Hafezi2013,He2016,Mukherjee2017}. The Hofstadter Hamiltonian 
for neutral lattice bosons in a synthetic magnetic fields has been 
experimentally implemented~\cite{Aidelsburger2013,Miyake2013}; with two spin
components, the system is time-reversal symmetric, yielding the neutral 
analog of the spin-Hall effect~\cite{Goldman2010}. The integer quantization of
the lowest-band Chern number was determined in the time-reversal-breaking 
geometry using transport measurements~\cite{Aidelsburger2015}. The topological 
Haldane model was realized by placing ultracold fermionic atoms in a 
periodically modulated optical honeycomb lattice~\cite{Jotsu2014}, and the 
Berry curvature was obtained using time-of-flight images of a Floquet 
lattice~\cite{Flaschner2016}. Most recently, a one-dimensional symmetry
protected topological phase was realized in an ultracold atomic 
gas~\cite{Song2017}.

Previous work has shown that particle dynamics can reveal the presence of 
topological order in systems that break time-reversal symmetry. Wave packets 
can acquire both anomalous velocities under applied forces~\cite{Price2012} and 
Berry-flux phases under closed trajectories in momentum space~\cite{Duca2015}.
The Berry curvature (whose integral over momentum space yields the Chern
number) can be obtained directly from time-of-flight 
images~\cite{Alba2011,Hauke2014}. Discrete-time quantum walks (i.e.\ dynamics
driven by a spin-dependent discrete-hopping model) have been shown to be 
affected by topology~\cite{Kitagawa2010,Obuse2011,Kitagawa2012,Asboth2012,Rakovszky2015,Asboth2015,Obuse2015,Cedzich2016}, and the moments of the
quantum walker probability distribution can be used as indicators of 
topological quantum phase transitions~\cite{Cardano2016}.

An on-going experimental challenge is the detection of topological order. In 
this work we show that the in-situ spin-dependent dynamics of 
particles driven by a two-dimensional spin-orbit Hamiltonian can indeed reveal 
both the presence of non-trivial topological order as well as the boundaries 
between different quantum phases. One need only prepare an initial localized 
state in the lattice and observe its density distribution under evolution. In 
the context of an ultracold atomic implementation, our results are robust 
against the localization of the initial state as well as the finite resolution 
of the optical imaging apparatus used to measure the particle distribution 
after some elapsed time. The results obtained in the present work are 
immediately applicable to on-going ultracold atom experiments~\cite{Wu2016}.

We consider the momentum-space Hamiltonian 
$H({\bm k})={\bm h}\cdot\boldsymbol\sigma$ for a two-component particle in two 
spatial dimensions, where
$\boldsymbol\sigma=(\sigma_x,\sigma_y,\sigma_z)$ is the three-vector of 
$2\times 2$ Pauli matrices and the components of ${\bm h}$ are each dependent 
on the quasimomenta ${\bm k}=(k_x,k_y)$. This spin-orbit interaction 
Hamiltonian, with momentum and spin degrees of freedom linked to each other, 
can support non-trivial topological phases. We employ the specific choice
\begin{eqnarray}
h_x&=&2t_1\cos(k_x);\quad h_y=2t_1\cos(k_y);\nonumber \\
h_z&=&m+2t_2\left[\sin(k_x)+\sin(k_y)\right],
\label{eq:kterms}
\end{eqnarray}
where $t_1$, $t_2$, and $m$ are adjustable parameters with units of energy;
this corresponds to a simplification of the model found in 
Ref.~\cite{Sticlet2012}. Most important, this is precisely the spin-orbit 
Hamiltonian recently realized experimentally with ultracold atoms in optical 
lattices~\cite{Wu2016}. In that work, experimental data were shown specifically 
for $t_2>t_1$ and $m\gg t_1$, but both $m/t_1$ and $t_2/t_1$ are adjustable 
over a wide range. The lattice momenta $k_x$ and $k_y$ are unitless as the 
adjustable lattice constant is assumed to be unity. The position space basis is 
the set of orthogonal states $c^{\dagger}_{i,j,\sigma}|{\mathcal O}\rangle
=|\sigma\rangle\otimes c^{\dagger}_{i,j}|{\mathcal O}\rangle
=|\sigma\rangle\otimes|i,j\rangle$,
where $\sigma=\{\uparrow,\downarrow\}$ and $|{\mathcal O}\rangle$ is the 
particle vacuum. The real-space complex lattice Hamiltonian giving rise to 
Eq.~(\ref{eq:kterms}) is then $H=H_0+H_{\rm hop}$, where 
\begin{eqnarray}
H_{\rm hop}&=\sum_{j,k}\left(t_xc_{j+1,k}^{\dagger}c_{j,k}
+t_yc_{j,k+1}^{\dagger}c_{j,k}+{\rm H.c.}\right)
\end{eqnarray}
corresponds to a particle hopping on a square lattice with complex 
spin-dependent amplitudes $t_x=t_1\sigma_x-it_2\sigma_z$ and 
$t_y=t_1\sigma_y-it_2\sigma_z$ along the $\hat{x}$ and $\hat{y}$ directions, 
and $H_0=\frac{m}{2}\sum_{j,k}\sigma_zc^{\dag}_{j,k}c_{j,k}$ is an on-site 
spin-dependent potential. This work employs periodic boundary conditions in 
both directions (two-torus geometry); for the analytical calculations we assume 
an infinite lattice but for the numerical results we necessarily employ a 
finite lattice.

The time-evolution of a state initially in spin up at the center of the 
two-dimensional lattice 
\begin{equation}
|\psi(0)\rangle=|\uparrow\rangle\otimes|0,0\rangle=
\left(\begin{matrix}
1\cr 0\cr
\end{matrix}\right)|0,0\rangle
\end{equation}
is most simply expressed in terms of the momentum-space eigenvalues 
$\pm E_{\bm k}=\pm\sqrt{h_x^2+h_y^2+h_z^2}$ and eigenvectors 
$|u_{n,{\bm k}}\rangle$ ($n=\pm$) of the two-band Hamiltonian as
\begin{eqnarray}
|\psi(t)\rangle
&=&\int\frac{d\bm k}{(2\pi)^2}\frac{h_x+ih_y}{2E_{\bm k}}
\left[-e^{iE_{\bm k}t/\hbar}
\begin{pmatrix}\frac{h_z-E_{\bm k}}{h_x+ih_y}\cr 1\end{pmatrix}\right.
\nonumber \\
&&\left.+e^{-iE_{\bm k}t/\hbar}\begin{pmatrix}\frac{h_z+E_{\bm k}}{h_x+ih_y}\cr 
1\end{pmatrix}\right]|{\bm k}\rangle.
\label{eq:psit}
\end{eqnarray}
It is evident that at $t=0$ the spatial wave function 
$\langle 0,0|\psi(0)\rangle$ is the uniform integral over all momentum states, 
as expected for a localized initial state. For finite times, the evolution over 
the spatial lattice mixes both eigenstates, allowing the particle to probe 
the full structure of both lower and upper bands. As shown below, this allows 
the dynamics to depend on topological features of the Hamiltonian.

\begin{figure}[t]
  \begin{center}
  \includegraphics[width=\columnwidth]{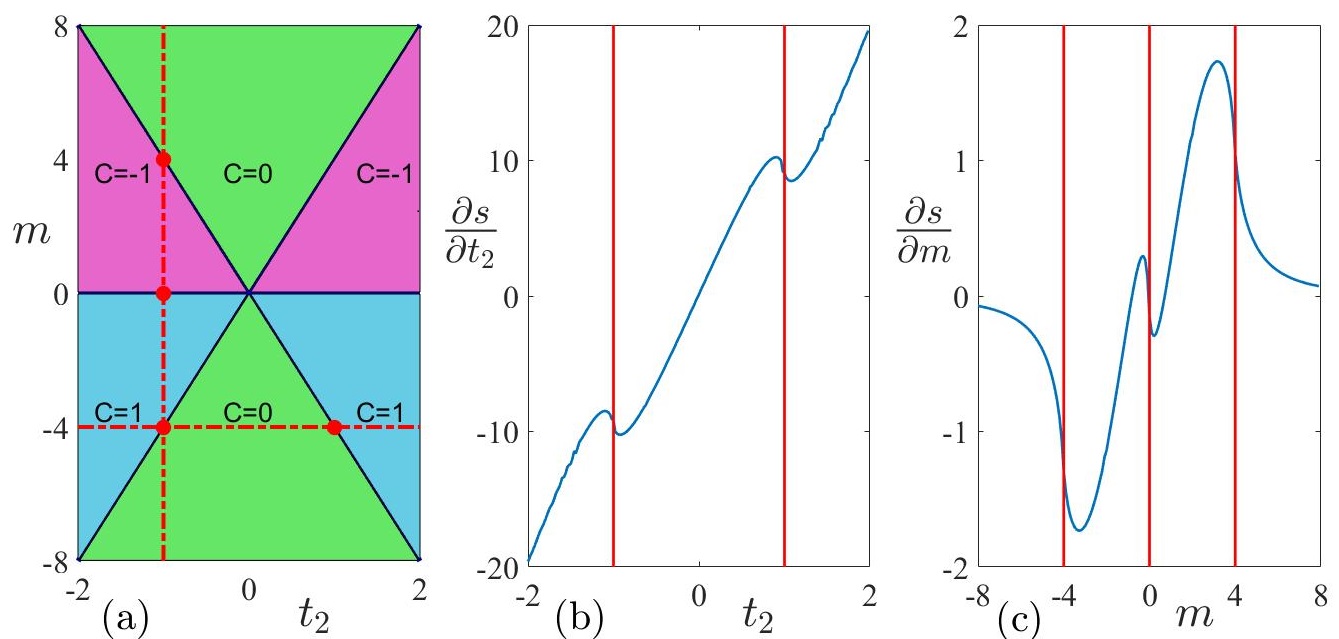}
  \end{center}
  \caption{The $m$--$t_2$ phase diagram for the spin-orbit Hamiltonian defined
by Eq.~(\ref{eq:kterms}) is shown in (a); different phases are characterized by 
the Chern numbers $C$. Red dots correspond to critical values of $t_2$ and $m$ 
along the $m=-4t_1$ and $t_2=-t_1$ dashed lines, respectively. The behaviors of
$\partial s/\partial t_2$ and $\partial s/\partial m$ along these directions 
are shown in (b) and (c) as a function of $t_2$ and $m$, respectively. Red 
vertical lines represent the phase boundaries, and we set $t=\hbar=t_1=1$.}
\label{fig:kinky}
\end{figure}

The topology of the Hamiltonian can be characterized by the gauge-dependent
Berry connections 
${\bm A}_{n'n}({\bm k})\equiv\langle u_{n',{\bm k}}|i\nabla_{\bm k}|
u_{n,{\bm k}}\rangle$ or the gauge-invariant Berry curvature 
$\Omega_{n'n}({\bm k})\equiv\nabla_{\bm k}\times{\bm A}_{n'n}({\bm k})$,
\begin{eqnarray}
\Omega({\bm k})_{n'n}&=&i\left\langle
\frac{\partial u_{n',{\bm k}}}{\partial k_x}\Bigg|
\frac{\partial u_{n,{\bm k}}}{\partial k_y}\right\rangle
-i\left\langle\frac{\partial u_{n',{\bm k}}}{\partial k_y}\Bigg|
\frac{\partial u_{n,{\bm k}}}{\partial k_x}\right\rangle.\hphantom{aaa}
\end{eqnarray}
In this work the relevant Chern number is defined as the topological invariant 
for the lower band $C\equiv(1/2\pi)\int d{\bm k}\;\Omega({\bm k})_{--}$; note 
that the sum of lower and upper-band Chern numbers is identically zero. For the
model~(\ref{eq:kterms}), one obtains after some straighforward algebra
\begin{equation}
C=\frac{t_1^2}{\pi}\int d\bm k\frac{2t_2\left(\sin k_x+\sin k_y\right)
+m\sin k_x\sin k_y}{E_{\bm k}^3}.
\label{eq:C}
\end{equation}
This integral that can be readily evaluated numerically, and one obtains 
$C=\{0,\pm 1\}$ depending on the 
choice of parameters $t_2$ and $m$ in units of $t_1$. The resulting phase 
diagram with regions characterized by different Chern numbers is shown in 
Fig.~\ref{fig:kinky}(a). The boundary between two topologically distinct phases
occurs when the gap between the two bands closes, i.e.\ at the Dirac points
$E_{\bm k}=0$ for some choice of parameters. Using the 
definitions~(\ref{eq:kterms}), the two bands touch at the pair of Dirac points
${\bm k}=\left(\pm\frac{\pi}{2},\mp\frac{\pi}{2}\right)$ when $m=0$ and at 
either of the single Dirac points 
${\bm k}=\left(\pm\frac{\pi}{2},\pm\frac{\pi}{2}\right)$ for the critical value 
of $t_2$
\begin{equation}
t_2^{c\pm}=\mp\frac{m}{4}.
\label{eq:t3c}
\end{equation}
These phase boundaries are shown in Fig.~\ref{fig:kinky}(a). 

For reasons that will become clear shortly, it is useful to consider the 
expression for the Chern number close to the phase transition. Choosing fixed
$m$ and $t_2=t_2^{c\pm}+\epsilon$, the leading contribution to the integrand of
Eq.~(\ref{eq:C}) comes from ${\bm k}$ values where $E_{\bm k}$ is minimized;
for sufficiently small $\epsilon$, these will be located in the vicinity of the 
Dirac points. Setting $t_2=t_2^{c\pm}+\epsilon$, 
$k_x\approx\pm\frac{\pi}{2}+k\cos(\phi)$ and 
$k_y\approx\pm\frac{\pi}{2}+k\sin(\phi)$, one finds that for $|m|/t_1\gg 1$
and $k\ll 1$, $E_{\bm k}^2$ is minimized for $k=k_c$, where 
\begin{equation}
k_c\approx 4\sqrt{\mp\epsilon/m}
\label{eq:kc}
\end{equation}
or $k_c=0$ if the above expression is imaginary. Setting $k=k_c$, one 
obtains $E_{\bm k}^2\approx 64t_1^2\left(\mp\epsilon/m\right)$ for 
$\mp\epsilon/m>0$ and $E_{\bm k}^2=16\epsilon^2$ for $\mp\epsilon/m<0$. To 
first order in $t_1/m$, the integrand of Eq.~(\ref{eq:C}) is highly peaked at 
$k=k_c$ and for $k\sim k_c$ is only weakly dependent on angle. 
Choosing for concreteness $t_2=t_2^{c+}+\epsilon$, one obtains
\begin{equation}
C\approx\frac{t_1^2}{2}\int_0^{\infty}k dk\frac{16\epsilon
-(m+4\epsilon)k^2}{\left(a+bk^2+ck^4\right)^{3/2}},
\label{eq:Capprox}
\end{equation}
where $a=16\epsilon^2$, $b=4t_1^2+2m\epsilon-8\epsilon^2$, and 
$c=m^2/16-t_1^2-5m\epsilon/8+3\epsilon^2/2$. For $m/t_1=-20$, this integral
can be readily evaluated numerically, yielding $C\approx 0.012$ and 
$C\approx 1.002$ for $\epsilon/t_1=-0.1$ and $0.1$, respectively. Similar 
results are obtained for $t_2=t_2^{c-}+\epsilon$.  
Thus, near the phase boundary for $|m|/t_1\gg 1$, the topological character is 
extremely well-captured by the lower band structure near the Dirac points.

Of particular interest to the present study is the spatial width of the 
particle distribution function. While this may be obtained directly from 
Eq.~(\ref{eq:psit}), a detailed calculation (found in the Supplementary 
Material) shows that at long times the experimentally observable quantity $s$, 
the time-derivative of the particle variance, depends only on the band 
structure
\begin{equation}
s\equiv\frac{\hbar^2}{t}\frac{\partial}{\partial t}\langle r^2\rangle
\approx\int\frac{d{\bm k}}{(2\pi)^2}\left(\nabla_{\bm k}E_{\bm k}\right)^2
=\int\frac{d{\bm k}}{(4\pi)^2}\frac{\left(\nabla_{\bm k}E_{\bm k}^2\right)^2}
{E_{\bm k}^2},
\label{eq:s}
\end{equation}
and not explicitly on the Berry connections or curvatures. Far from the phase 
boundary $|t_2|\gg\left|t_2^{c\pm}\right|$, one can evaluate Eq.~(\ref{eq:s}) 
analytically in the limit $|m|/t_1\gg 1$. One obtains $s=4t_2^2$ and therefore 
$\partial s/\partial t_2=8t_2$, whose linear dependence on $t_2$ is confirmed 
by the numerical results shown in Fig.~\ref{fig:kinky}(b). Likewise, it is 
straightforward to show that $\partial s/\partial m=0$ for $|m|\gg t_1$, 
consistent with the edges of Fig.~\ref{fig:kinky}(c).

Pronounced `kinks' in the variations of $\partial s/\partial t_2$ with $t_2$
and of $\partial s/\partial m$ with $m$ can also be seen in 
Figs.~\ref{fig:kinky}(b) and (c), respectively, clearly revealing the quantum 
phase transitions. Consider the variation of $\partial s/\partial t_2$ with 
$t_2$ (the other case $t_2\leftrightarrow m$ proceeds analogously). Close to 
the phase boundary $t_2=t_2^{c\pm}+\epsilon$ and considering only 
$\mp\epsilon/m>0$ (i.e.\ $C=\pm 1$), one obtains
\begin{equation}
s\approx\frac{512t_1^4}{\pi}\left(\mp\frac{\epsilon}{m}\right)
\int_0^{\infty}\frac{k dk}{a+bk^2+ck^4},
\label{eq:sapprox}
\end{equation}
where $a$, $b$, and $c$ take the same values as in Eq.~(\ref{eq:Capprox}).
Defining $k^2\equiv\sqrt{a/c}x$ the integral is readily evaluated
analytically, yielding
\begin{equation}
s\approx\frac{512t_1^4}{\pi}\left(\mp\frac{\epsilon}{m}\right)
\frac{\cos^{-1}\left(b/2\sqrt{ac}\right)}{\sqrt{4ac-b^2}}.
\end{equation}
Again for $|m|/t_1\gg 1$, one has
\begin{equation}
s\approx -\frac{256t_1^4}{\pi m^2}\left(1-\frac{2t_1^2}{3m\epsilon}
\right).
\end{equation}
This gives $\partial s/\partial t_2\sim\partial s/\partial\epsilon\sim
(t_1^6/|m|^3)(t_2-t_2^{c\pm})^{-2}$, which strongly deviates from the linear
dependence on $t_2$ far from the phase boundary. In fact, the slope of this
function is negative, as confirmed by the numerical data presented in
Fig.~\ref{fig:kinky}(b).

The deviation from the linear $t_2$-dependence of $\partial s/\partial t_2$ 
near the phase transition becomes increasingly pronounced as $m$ decreases, 
which should aid in its experimental detection. Likewise, the signature of the
phase transition becomes stronger for higher-order moments 
$\partial^{k}s/\partial t_2^{k}$, though these would likely be difficult 
to measure precisely in experiments. Though the analytics above only considered 
the behavior in one phase, the numerical results depicted in 
Fig.~\ref{fig:kinky}(b) and (c) clearly show a similar kink near the boundary 
for all phases, and one may infer similar behavior for crossings not shown. We 
have verified similar behavior for the triangular lattice~\cite{Sticlet2012} 
which supports states with $C=2$. These findings are consistent with those 
obtained for one-dimensional 
systems~\cite{Cardano2016}. We claim that the numerical and analytical results 
provide clear evidence that the energy band gap closes; while this is necessary 
between different topologically ordered phases or between a trivial and 
non-trivial phase, it is not sufficient to indicate topological order as gap 
closure could occur between two trivial phases. As such, we will provide 
additional evidence supporting the topological nature of the phase transition.

\begin{figure}[t]
\includegraphics [width=\columnwidth]{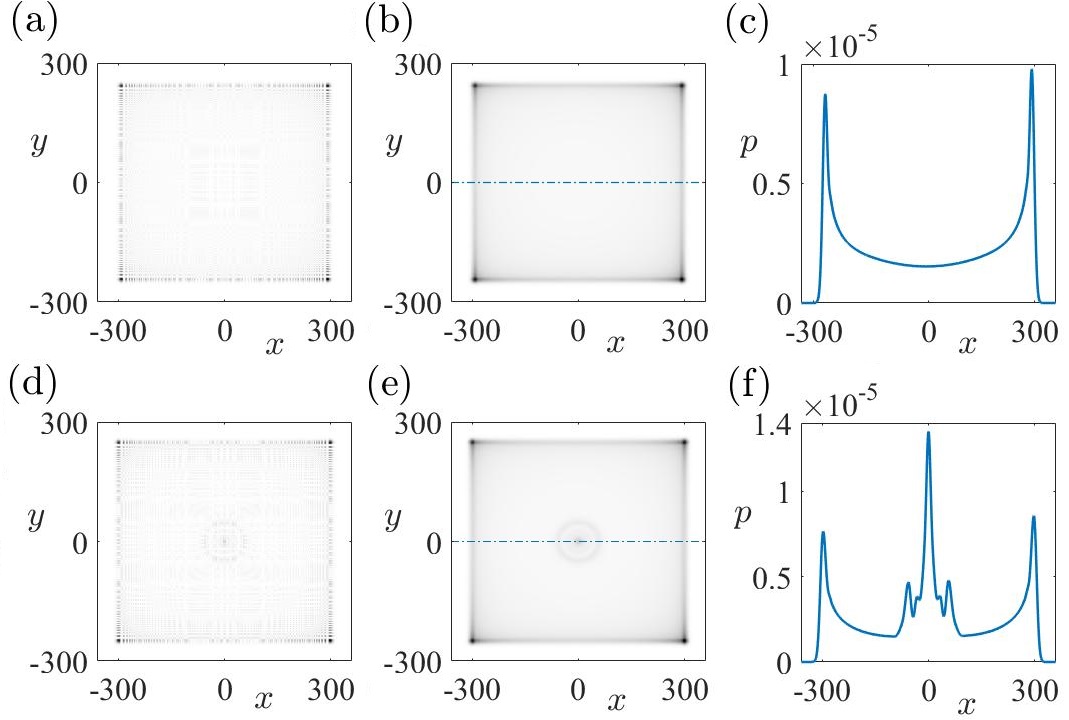}
\caption{Particle density distributions $p=\langle{\bm r}|\psi(t)\rangle|^2$.
Results for $m=-20t_1$ and a $599\times 599$ lattice are shown in (a)-(c) for 
$t_2=4.5t_1$ and $t=28\hbar/t_1$ ($C=0$), and (d)-(f) for $t_2=5.5t_1$ and 
$t=23\hbar/t_1$ ($C=1$). 
Raw data are presented in (a) and (d); 
smoothed densities in (b) and (e) are obtained by convolving with the function 
$e^{-(k_x^2+k_y^2)/\gamma^2}$, $\gamma=12.45$, representing the finite 
resolution of experimental imaging and initial state preparation. Slices
through the centers of the particle densities (blue dashed lines) are shown in 
(c) and (f).}
\label{fig:densities}
\end{figure}

With the appropriate parameter choices, the {\it in situ} density profile
itself can reveal the nature of the quantum phase, which the average width of
the particle distribution cannot. Fig.~\ref{fig:densities} shows characteristic 
snapshots of the time-evolved real-space probability 
$|\langle{\bm r}|\psi(t)\rangle|^2$ when $|m|/t_1\gg 1$, obtained from 
applying a discrete Fourier transform to $\langle{\bf k}|\psi(t)\rangle$ at 
long times, subject to ensuring that the leading front of the particle density 
remains negligible at the edge of the physical lattice. In the trivial phase 
characterized by $C=0$, Figs.~\ref{fig:densities}(a)-(c), the density is 
generally featureless with a maximum in the vicinity of the leading edge.
This is clearly visible in the slice through the density center, 
Fig.~\ref{fig:densities}(c) and dovetails with results for a discrete-time 
quantum walk on a square lattice~\cite{Watabe2008}. For the free lattice 
evolution investigated in this 
work (equivalent to a continuous-time quantum walk) the density profile at long 
times is square. This result is consistent with experiments on ultracold atoms 
expanding in a square optical lattice in the absence of particle 
interactions~\cite{Schneider2012}, which show little influence on the extent of 
the initial particle localization. Indeed, choosing a less localized initial 
state in the numerics, such as a Gaussian distribution with different widths, 
leads to similar final densities for sufficiently long evolution times (not 
shown). Likewise, the results hold if the finite resolution of the experimental 
imaging system and the initial state preparation is taken into account by 
convolving the particle distribution with a Gaussian 
$e^{-(k_x^2+k_y^2)/\gamma^2}$, $\gamma=12.45$, representing a generic 
point-spread function; corresponding results are shown in 
Fig.~\ref{fig:densities}(b).

In the topologically non-trivial phase with $C=\pm 1$, the time-evolved density 
profile is similar to that in the trivial phase, but reveals additional rings 
of high density in the neighborhood of the lattice centre where the particle 
originated, as shown in Fig.~\ref{fig:densities}(d)-(f). These rings again 
remain well-defined under changes in initial conditions or under the smoothing 
due to the finite imaging resolution, Fig.~\ref{fig:densities}(e). While the 
central peak is clearly visible in the density plots \ref{fig:densities}(d) and 
(e), the fainter adjacent ring is more pronounced in the slice through the 
center, Fig.~\ref{fig:densities}(f). The ring profile is independent of time; 
the peak positions remain essentially fixed even as the density as a whole 
expands.

The appearance of these peaks in the density is tied closely to the underlying 
topology. Because the rotation of the particle spin is tied to its momentum, 
the time-evolved state $|\psi(t)\rangle$, Eq.~(\ref{eq:psit}), becomes a 
superposition of spin-up and spin-down states. 
In principle, these can be independently imaged experimentally. Consider first 
the spin-down component which is initially unpopulated; close to the phase 
boundary and near the particle origin, the real-space wavefunction is 
approximately the cylindrical Fourier transform
\begin{eqnarray}
\psi_{\downarrow}(r,t)&\approx&\frac{it_1}{2\pi^2}\sin(4\epsilon t/\hbar)
\int\frac{k dk d\phi e^{-i\left(\frac{\pi}{2}+k\right)r\cos\phi}
k e^{i\phi}}{\sqrt{a+bk^2+ck^4}}\nonumber \\
&=&\frac{t_1}{\pi}\sin(4\epsilon t/\hbar)\int_0^{\infty}
\frac{k^2dk J_1\left[r\left(\frac{\pi}{2}+k\right)\right]}
{\sqrt{a+bk^2+ck^4}},
\end{eqnarray}
with the parameters $a$, $b$, and $c$ again the same as those in 
Eq.~(\ref{eq:Capprox}) and $J_1(x)$ is the Bessel function of the first kind. 
For $C=\pm 1$, the integrand is dominated by the contribution $k=k_c$, 
Eq.~(\ref{eq:kc}). To capture the qualitative properties assume that only the 
$k=k_c$ term contributes; one then obtains
\begin{equation}
\psi_{\downarrow}(r,t)\approx\mp\frac{4t_1}{m\pi}\sin\left(
\frac{4\epsilon t}{\hbar}\right)J_1\left(\frac{\pi r}{2}\right).
\end{equation}
The spin-down density for $C=\pm 1$ (for parameters close to the phase 
boundary) therefore displays a series of concentric rings in the vicinity of 
the particle origin, spaced by the maxima of the Bessel function. It is 
straightforward to show that a similar result also holds for the spin-up 
density, except with the maxima of $J_0$ (so that the first maximum is at the 
origin rather than slightly displaced). In contrast, for $C=0$ all values of 
$k$ contribute to the integral. The integrand therefore consists of a sum of
$J_{\alpha}(kr)$ factors with different values of $k$, which has the effect of
smearing out the Bessel function maxima; thus, for $C=0$ the concentric rings
are not manifested.

The ring of peaks in the particle distribution appear only in the non-trivial
topological phase, but not immediately at the phase boundary. To determine the
parameters, it suffices to calculate maxima of the lower band 
$\nabla_{\bm k}E_{\bm k}=0$ for $k_x=k_y$. Simple algebra yields solutions
$k_x=\left\{\pm\frac{\pi}{2},\sin^{-1}\left[\frac{m(m-4\epsilon)}
{(m-4\epsilon)^2-8t_1^2}\right]\right\}$, i.e.\ the two Dirac points and an 
additional ring. The last solution is real only if the argument is unity or 
smaller, which gives $\epsilon\geq -2t_1^2/m$. Thus, for $|m|/t_1\gg 1$, the 
central peaks manifest themselves almost immediately upon crossing into the 
non-trivial phase, while for smaller $|m|/t_1$ they appear deeper in the phase. 
These smaller (larger) values of $m$ ($\epsilon$) invalidate the analytical 
approximations made above, for example neglecting the angular dependence of
the energy minima as in Eq.~(\ref{eq:kc}), and consequently the central peaks 
would not be as clearly observable. In practice, the ring features become 
increasingly washed out for $|m|\lesssim 3t_1$ and $|\epsilon|\gtrsim t_1$. A 
similar effect would likely occur for Hamiltonians with closely-spaced Dirac 
points.

We now argue that the peaks in the particle density obtained above are generic,
subject to some restrictions on the choice of Hamiltonian. All of the results 
hinge on the appearance in the $C\neq 0$ phase of a ring of energy minima at 
momenta distributed at a radius $k_c$ from the Dirac point. Following 
for instance \cite{Sticlet2012}, one can consider ${\bm h}({\bm k})
=(h_x,h_y,h_z)$ as
a closed two-dimensional parametric surface ${\mathcal M}$, with the 
Dirac points defined by the origin $(0,0,0)$. The Chern number is then defined 
as the number of times this oriented surface wraps around the origin; if 
${\mathcal M}$ touches the origin the band gap closes and $C$ is not defined.

The Hamiltonian~(\ref{eq:kterms}) belongs to a family of the form 
$H({\bm k})=(h_x,h_y,m+t_2g_z)\cdot\boldsymbol\sigma$, where $h_x$, $h_y$ and 
$g_z$ are 
periodic, non-constant functions, symmetric around the $\sigma_z$-axis and such 
that the parametric surface ${\mathcal M}$ has a negative-definite curvature. 
In the large-mass limit $|m|\gg t_1$, a topological phase transition is 
generically characterized by a ring in the energy surface. The phase transition 
occurs for large $t_2^c\sim|m|$ from Eq.~(\ref{eq:t3c}), stretching the surface 
along the $\sigma_z$-axis. The outer points of ${\mathcal M}$ are approximately 
distributed on the surface of a prolate spheroid which passes through the 
origin as $t_2$ increases, changing the Chern number from zero to $\pm 1$. 
On the trivial side, the energy surface will show a single minimum, 
corresponding to the outer point of the spheroid. After passing through, a ring 
of minima appears, symmetric around the $\sigma_z$-axis.

The numerical and analytical results presented here indicate that for a
realistic (experimentally motivated) spin-orbit lattice Hamiltonian, it is 
possible to detect transitions between topological phases by allowing particles
to evolve freely in the lattice and observe their spatial distribution. The 
presence of topological order can be inferred from the onset of spatial peaks
in the vicinity of the particle origin, which remain well-defined even taking
into consideration finite imaging resolution. Similar results are also found 
for particles hopping on a triangular lattice. This technique is readily 
applicable to recent ultracold atom experiments~\cite{Wu2016} and to future 
implementations using photonic lattices, and should aid in the detection of 
topological transitions in these systems.

\begin{acknowledgements}
The authors would like to acknowledge useful discussions with Shuai Chen and 
Wei Sun from USTC. This research was supported by NSERC (Canada), the China 
1000 Talent Plan, the National Natural Science Foundation of China (Grant No.\ 
GG2340000241), the China Scholarship Council, and the Australian Research 
Council (project number CE110001013).
\end{acknowledgements}

\bibliographystyle{aip}
\bibliography{topo}

\appendix

\begin{widetext}

\begin{center}
{\bf Supplementary Material}
\end{center}

\m Consider a generic (two-dimensional) spin-orbit Hamiltonian. It can be written 
as
\begin{equation}
H(k_x,k_y)=\begin{pmatrix}
h_z & h_x-ih_y\cr
h_x+ih_y & -h_z\end{pmatrix},
\end{equation}
where $h_i=h_i(k_x,k_y)$. This matrix has eigenvalues
$E_{\pm}(k_x,k_y)=E_{\pm}({\bm k})=\pm\sqrt{h_x^2+h_y^2+h_z^2}
\equiv\pm E_{\bm k}$, and associated
eigenvectors
\begin{equation}
|u_-\rangle=\sqrt{E_{\bm k}+h_z\over 2E_{\bm k}}\begin{pmatrix}\frac{h_z-E_{\bm k}}{h_x+ih_y}\cr 1\end{pmatrix};
\qquad
|u_+\rangle=\sqrt{E_{\bm k}-h_z\over 2E_{\bm k}}\begin{pmatrix}\frac{h_z+E_{\bm k}}{h_x+ih_y}\cr 1\end{pmatrix}.
\label{eq:Blochfunctions}
\end{equation}
The initial state corresponds to a particle localized to a single lattice
point. This lattice point $(0,0)$ has the simplest expression in Fourier
coordinates:
\begin{equation}
|0,0\rangle_{x,y}=\sum_{k_x,k_y}|k_x,k_y\rangle,
\end{equation}
so that the initial state is an equal superposition of every state in 
${\bm k}$-space. In addition, suppose the particle starts in spin up:
\begin{equation}
\begin{pmatrix}1\cr 0\end{pmatrix}
=\frac{h_x+ih_y}{\sqrt{2E_{\bm k}}}\left(-\frac{1}{\sqrt{E_{\bm k}+h_z}}|u_-\rangle
+\frac{1}{\sqrt{E_{\bm k}-h_z}}|u_+\rangle\right).
\label{eq:spinstart}
\end{equation}
The (spinor) initial state is therefore
\begin{eqnarray}
|\psi(0)\rangle\equiv\begin{pmatrix}1\cr 0\end{pmatrix}|0,0\rangle_{x,y}
&=&\sum_{k_x,k_y}\frac{h_x+ih_y}{\sqrt{2E_{\bm k}}}\left(
-\frac{1}{\sqrt{E_{\bm k}+h_z}}\langle k_x,k_y|u_-\rangle+\frac{1}{\sqrt{E_{\bm k}-h_z}}
\langle k_x,k_y|u_+\rangle\right)|k_x,k_y\rangle\nonumber \\
&=&\sum_{k_x,k_y}\frac{h_x+ih_y}{2E_{\bm k}}\left[
-\begin{pmatrix}\frac{h_z-E_{\bm k}}{h_x+ih_y}\cr 1\end{pmatrix}+\begin{pmatrix}\frac{h_z+E_{\bm k}}{h_x+ih_y}\cr 1\end{pmatrix}
\right]|k_x,k_y\rangle=\sum_{k_x,k_y}\begin{pmatrix}1\cr 0\end{pmatrix}
|k_x,k_y\rangle,\hphantom{aa}
\end{eqnarray}
as expected. Importantly, the initial state is a superposition of both lower 
($n=-$) and upper ($n=+$) bands, so the evolution of the position variance with 
time requires a full mixing of both bands: 
\begin{equation}
|\psi(t)\rangle
=\sum_{\bm k}\frac{h_x+ih_y}{2E_{\bm k}}\left[-e^{iE_{\bm k}t/\hbar}
\begin{pmatrix}\frac{h_z-E_{\bm k}}{h_x+ih_y}\cr 1\end{pmatrix}
+e^{-iE_{\bm k}t/\hbar}\begin{pmatrix}\frac{h_z+E_{\bm k}}{h_x+ih_y}\cr 
1\end{pmatrix}\right]|{\bm k}\rangle.
\end{equation}
Also, the
expressions for the $u_{\pm}(k_x,k_y)\equiv\langle k_x,k_y|u_{\pm}\rangle$ are
non-separable, i.e.\ one cannot write $u_-(k_x,k_y)=u_-(k_x)u_-(k_y)$ and 
similarly for $u_+(k_x,k_y)$.

\mm  The expressions for the Bloch functions~(\ref{eq:Blochfunctions}) allow us 
to calculate the Berry connections. After some algebraic manipulations one 
obtains
\begin{eqnarray}
A_x({\bm k})_{--}&\equiv&\left\langle u_-\bigg|i\frac{\partial}{\partial k_x}
\bigg|u_-\right\rangle=\frac{1}{2E_{\bm k}(E_{\bm k}+h_z)}\left(h_x
\frac{\partial h_y}{\partial k_x}-h_y\frac{\partial h_x}{\partial k_x}\right);
\nonumber \\
A_x({\bm k})_{++}&\equiv&\left\langle u_+\bigg|i\frac{\partial}{\partial k_x}
\bigg|u_+\right\rangle=\frac{1}{2E_{\bm k}(E_{\bm k}-h_z)}\left(h_x
\frac{\partial h_y}{\partial k_x}-h_y\frac{\partial h_x}{\partial k_x}\right).
\end{eqnarray}
The expressions for $A_y$ are obtained analogously, by replacing the
$k_x$-derivatives by $k_y$-derivatives. Note that these intra-band Berry
connections are purely real, as expected from the fact that
$\langle u_n|i\partial/\partial k_i|u_n\rangle^{\dag}
=\langle u_n|i\partial/\partial k_i|u_n\rangle$ (the real $x$ operator maps to
$i\partial/\partial k_x$). The corresponding inter-band Berry connections are
\begin{eqnarray}
A_x({\bm k})_{+-}&\equiv&\left\langle u_+\bigg|i\frac{\partial}{\partial k_x}
\bigg|u_-\right\rangle=-\frac{1}{2E_{\bm k}\sqrt{h_x^2+h_y^2}}\left[
h_x\frac{\partial h_y}{\partial k_x}-h_y\frac{\partial h_x}{\partial k_x}
+i\left(h_z\frac{\partial E_{\bm k}}{\partial k_x}-E_{\bm k}\frac{\partial h_z}{\partial k_x}
\right)\right];\nonumber \\
A_x({\bm k})_{-+}&\equiv&\left\langle u_-\bigg|i\frac{\partial}{\partial k_x}
\bigg|u_+\right\rangle=-\frac{1}{2E_{\bm k}\sqrt{h_x^2+h_y^2}}\left[
h_x\frac{\partial h_y}{\partial k_x}-h_y\frac{\partial h_x}{\partial k_x}
-i\left(h_z\frac{\partial E_{\bm k}}{\partial k_x}-E_{\bm k}\frac{\partial h_z}{\partial k_x}
\right)\right].
\end{eqnarray}
Again, the expressions for $A_y({\bm k})_{nn'}$ are obtained by replacing
$k_x$-derivatives with $k_y$-derivatives. The inter-band Berry connections
properly satisfy the expected relationship $A_i({\bm k})_{nn'}^*
=A_i({\bm k})_{n'n}$, $n\neq n'$. Surprisingly, however, they are complex
quantities. That said, the real parts are proportional to their intra-band
counterparts. The Berry curvature is defined as
\begin{equation}
\Omega({\bm k})_{n'n}=\left[\nabla_k\times{\bm A}({\bm k})
\right]_{n'n}=\frac{\partial}{\partial k_x}A_y({\bm k})_{n'n}
-\frac{\partial}{\partial k_y}A_x({\bm k})_{n'n}.
\label{eq:Berrycurvature}
\end{equation}
After some algebra, the Berry curvature of interest takes the simple and
intuitive form:
\begin{equation}
\Omega({\bm k})_{--}=\frac{1}{2E_{\bm k}^3}\left[
h_x\left(
\frac{\partial h_y}{\partial k_x}\frac{\partial h_z}{\partial k_y}
-\frac{\partial h_y}{\partial k_y}\frac{\partial h_z}{\partial k_x}
\right)
+h_y\left(
\frac{\partial h_z}{\partial k_x}\frac{\partial h_x}{\partial k_y}
-\frac{\partial h_z}{\partial k_y}\frac{\partial h_x}{\partial k_x}
\right)
+h_z\left(
\frac{\partial h_x}{\partial k_x}\frac{\partial h_y}{\partial k_y}
-\frac{\partial h_x}{\partial k_y}\frac{\partial h_y}{\partial k_x}
\right)\right].
\end{equation}

Let's consider the second term of Eq.~(\ref{eq:Berrycurvature}):
\begin{eqnarray}
-\frac{\partial}{\partial k_y}A_x({\bm k})_{n'n}&=&-\sum_{\sigma}\left[
\left(\frac{\partial}{\partial k_y}u^*_{n',{\bm k},\sigma}\right)i
\frac{\partial}{\partial k_x}u_{n,{\bm k},\sigma}
+u^*_{n',{\bm k},\sigma}i\frac{\partial^2}{\partial k_xk_y}
u_{n,{\bm k},\sigma}\right)
\nonumber \\
&=&-i\left\langle\frac{\partial}{\partial k_y}u_{n',{\bm k}}\Bigg|
\frac{\partial}{\partial k_x}u_{n,{\bm k}}\right\rangle
-i\left\langle u_{n',{\bm k}}\Bigg|\frac{\partial^2}{\partial k_yk_x}
u_{n,{\bm k}}\right\rangle.
\end{eqnarray}
The first term follows by interchanging $k_x\leftrightarrow k_y$:
\begin{eqnarray}
\frac{\partial}{\partial k_x}A_y({\bm k})_{n'n}&=&\sum_{\sigma}\left[
\left(\frac{\partial}{\partial k_x}u^*_{n',{\bm k},\sigma}\right)i
\frac{\partial}{\partial k_y}u_{n,{\bm k},\sigma}
+u^*_{n',{\bm k},\sigma}i\frac{\partial^2}{\partial k_xk_y}
u_{n,{\bm k},\sigma}\right)\nonumber \\
&=&i\left\langle\frac{\partial}{\partial k_x}u_{n',{\bm k}}\Bigg|
\frac{\partial}{\partial k_y}u_{n,{\bm k}}\right\rangle
+i\left\langle u_{n',{\bm k}}\Bigg|\frac{\partial^2}{\partial k_xk_y}
u_{n,{\bm k}}\right\rangle.
\end{eqnarray}
Because $\frac{\partial^2}{\partial k_xk_y}=\frac{\partial^2}{\partial k_yk_x}$
one obtains
\begin{equation}
\Omega({\bm k})_{z,n'n}=i\left\langle\frac{\partial}{\partial k_x}
u_{n',{\bm k}}\Bigg|\frac{\partial}{\partial k_y}u_{n,{\bm k}}\right\rangle
-i\left\langle\frac{\partial}{\partial k_y}u_{n',{\bm k}}\Bigg|
\frac{\partial}{\partial k_x}u_{n,{\bm k}}\right\rangle,
\end{equation}
which is the well-known expression.

\mm  We would like to calculate the time-dependent expectation value of
\begin{equation}
\langle r^2\rangle=\langle x^2+y^2\rangle
=\langle\psi(t)|x^2+y^2|\psi(t)\rangle=\sum_{{\bm r},{\bm r}'}\langle\psi(t)|
{\bm r}\rangle\langle{\bm r}|x^2+y^2|{\bm r}'\rangle\langle{\bm r}'|\psi(t)
\rangle=\sum_{{\bm r},\sigma}\psi_{\sigma}^*({\bm r},t)(x^2+y^2)\psi_{\sigma}
({\bm r},t),
\label{eq:r2exp}
\end{equation}
to find out the dependence on the Berry curvature (if any). Here the sum over 
$\sigma$ corresponds to adding the two components $-$ and $+$ of the two-band 
state. The state $\psi_{\sigma}({\bm r},t)$ can be expressed as
\begin{equation}
\psi_{\sigma}({\bm r},t)=\sum_n\sum_{{\bm k}}a_n({\bm k},t)
\psi_{n,{\bm k},\sigma}({\bm r}),
\label{eq:psidef1}
\end{equation}
where the arbitrary state is expanded in a complete basis set,
\begin{equation}
\sum_{{\bm r},\sigma}\psi^*_{n',{\bm k}',\sigma}({\bm r})
\psi_{n,{\bm k},\sigma}({\bm r})=\delta_{n,n'}\delta({\bm k}'-{\bm k}).
\label{eq:ortho}
\end{equation}
Note that the states $\psi_{n,{\bm k},\sigma}({\bm r})$ are the components of
the solutions of the full Hamiltonian on the lattice,
\begin{equation}
H|\psi_{n,{\bm k}}\rangle=E_n({\bm k})|\psi_{n,{\bm k}}\rangle,
\end{equation}
where $\psi_{n,{\bm k},\sigma}({\bm r})
=\langle{\bm r},\sigma|\psi_{n,{\bm k}}\rangle$. These
basis function are not translationally invariant because according to Bloch's
theorem
\begin{equation}
\psi_{n,{\bm k},\sigma}({\bm r}+{\bm R})=e^{i{\bm k}\cdot{\bm R}}
\psi_{n,{\bm k},\sigma}({\bm r}),
\label{eq:nottrans}
\end{equation}
where ${\bm R}$ is an arbitrary unit cell lattice vector. So it is more
convenient to expand in translationally invariant Bloch functions where
$\psi_{n,{\bm k},\sigma}({\bm r})\equiv e^{i{\bm k}\cdot{\bm r}}
u_{n,{\bm k},\sigma}({\bm r})$, where the functions satisfy
$u_{n,{\bm k},\sigma}({\bm r}+{\bm R})=u_{n,{\bm k},\sigma}({\bm r})$. It's
clear that
this definition is consistent with Eq.~(\ref{eq:nottrans}). Also because these
functions have the same value for any ${\bm r}$ we can drop the
${\bm r}$-dependence completely. Note that these basis functions are the
eigenfunctions of the ${\bm k}$-dependent Hamiltonian
$H({\bm k})\equiv e^{-i{\bm k}\cdot{\bm r}}He^{i{\bm k}\cdot{\bm r}}$:
\begin{equation}
H({\bm k})|u_{n,{\bm k}}\rangle=E_n({\bm k})|u_{n,{\bm k}}\rangle.
\label{eq:BlochHam}
\end{equation}
The definition of the arbitrary function, Eq.~(\ref{eq:psidef1}), then becomes
\begin{equation}
\psi_{\sigma}({\bm r},t)=\sum_n\sum_{\bm k}a_n({\bm k},t)
e^{i{\bm k}\cdot{\bm r}}u_{n,{\bm k},\sigma}.
\label{eq:psidef2}
\end{equation}
Comparison of Eqs.~(\ref{eq:spinstart}) and (\ref{eq:psidef2}) (recall that
$|{\bm r}|=0$) gives
\begin{equation}
a_-(k_x,k_y)=-\frac{h_x+ih_y}{\sqrt{2E_{\bm k}(E_{\bm k}+h_z)}};\quad
a_+(k_x,k_y)=\frac{h_x+ih_y}{\sqrt{2E_{\bm k}(E_{\bm k}-h_z)}}.
\end{equation}
for the particular problem of interest.
The orthogonality condition~(\ref{eq:ortho}) then becomes
\begin{equation}
\sum_{{\bm r},\sigma}e^{i({\bm k}-{\bm k}')\cdot{\bm r}}
u^*_{n',{\bm k}',\sigma}({\bm r})u_{n,{\bm k},\sigma}({\bm r})
=\delta_{n,n'}\delta({\bm k}'-{\bm k}).
\end{equation}
But on a lattice the $u_{n,{\bm k},\sigma}({\bm r})$ have the same value for
any ${\bm r}$; the ${\bm r}$-dependence can therefore be dropped, and one
obtains
\begin{equation}
\sum_{{\bm r},\sigma}e^{i({\bm k}-{\bm k}')\cdot{\bm r}}
u^*_{n',{\bm k}',\sigma}u_{n,{\bm k},\sigma}
=\langle u_{n',{\bm k}'}|u_{n,{\bm k}}\rangle
\sum_{{\bm r}}e^{i({\bm k}-{\bm k}')\cdot{\bm r}}
=\langle u_{n',{\bm k}'}|u_{n,{\bm k}}\rangle\delta({\bm k}-{\bm k}')
=\delta_{n,n'}\delta({\bm k}'-{\bm k}).
\label{eq:ortho2}
\end{equation}
Eq.~(\ref{eq:r2exp}) is then written
\begin{eqnarray}
\left\langle x^2+y^2\right\rangle
&=&\sum_{{\bm r},\sigma}\sum_{n,n'}\sum_{\bm k}\sum_{{\bm k}'}
a^*_{n'}({\bm k}',t)\psi^*_{n',{\bm k}',\sigma}({\bm r})(x^2+y^2)a_n({\bm k},t)
\psi_{n,{\bm k},\sigma}({\bm r})\nonumber \\
&=&\sum_{{\bm r},\sigma}\sum_{n,n'}\sum_{\bm k}\sum_{{\bm k}'}
a^*_{n'}({\bm k}',t)a_n({\bm k},t)e^{-i{\bm k}'\cdot{\bm r}}
u^*_{n',{\bm k}',\sigma}(x^2+y^2)e^{i{\bm k}\cdot{\bm r}}u_{n,{\bm k},\sigma}.
\label{eq:r2exp2}
\end{eqnarray}
It is convenient to focus on one dimension at a time.
\begin{eqnarray}
\left\langle x^2\right\rangle&=&
\sum_{x,\sigma}\sum_{n,n'}\sum_{\bm k}\sum_{{\bm k}'}
a^*_{n'}({\bm k}',t)a_n({\bm k},t)e^{-ik_x'x}u^*_{n',{\bm k}',\sigma}x^2
e^{ik_xx}u_{n,{\bm k},\sigma}\sum_{y}e^{i(k_y-k_y')y}\nonumber \\
&=&
\sum_{x,\sigma}\sum_{n,n'}\sum_{\bm k}\sum_{{\bm k}'}
a^*_{n'}({\bm k}',t)a_n({\bm k},t)e^{-ik_x'x}u^*_{n',{\bm k}',\sigma}x^2
e^{ik_xx}u_{n,{\bm k},\sigma}\delta(k_y-k_y')\nonumber \\
&=&\sum_{n,n'}\sum_{k_x,k_x'}\sum_{k_y}a^*_{n'}(k_x',k_y,t)
a_n(k_x,k_y,t)\sum_{x,\sigma}e^{-ik_x'x}u^*_{n',k_x',k_y,\sigma}x^2e^{ik_xx}
u_{n,k_x,k_y,\sigma}.
\label{eq:x2exp}
\end{eqnarray}
Note that we could equivalently express the final sum as
\begin{equation}
\sum_{x,\sigma}\psi^*_{n',k_x',k_y,\sigma}(x)x^2\psi_{n,k_x,k_y,\sigma}(x)
=\sum_{x,x'}\langle\psi_{n',k_x',k_y}|x\rangle\langle x|x^2|x'\rangle\langle x'|
\psi_{n,k_x,k_y}\rangle=\langle\psi_{n',k_x',k_y}|x^2|\psi_{n,k_x,k_y}\rangle.
\end{equation}

\mm  To proceed, note that:
\begin{eqnarray}
\frac{\partial^2}{\partial k_x^2}\sum_{x,\sigma}\psi^*_{n',k_x',k_y,\sigma}(x)
\psi_{n,k_x,k_y,\sigma}(x)
&=&\frac{\partial^2}{\partial k_x^2}\sum_{x,\sigma}e^{-ik_x'x}
u^*_{n',k_x',k_y,\sigma}e^{ik_xx}u_{n,k_x,k_y}\nonumber \\
&=&\sum_{x,\sigma}e^{-ik_x'x}u^*_{n',k_x',k_y,\sigma}
\frac{\partial^2}{\partial k_x^2}\left[e^{ik_xx}u_{n,k_x,k_y,\sigma}\right]
\nonumber \\
&=&\sum_{x,\sigma}e^{-ik_x'x}u^*_{n',k_x',k_y,\sigma}
\frac{\partial}{\partial k_x}\left[ixe^{ik_xx}u_{n,k_x,k_y,\sigma}
+e^{ik_xx}\frac{\partial}{\partial k_x}u_{n,k_x,k_y,\sigma}\right]\nonumber \\
&=&\sum_{x,\sigma}e^{-ik_x'x}u^*_{n',k_x',k_y,\sigma}\left[-x^2e^{ik_xx}
+2ixe^{ik_xx}\frac{\partial}{\partial k_x}+e^{ik_xx}
\frac{\partial^2}{\partial k_x^2}\right]u_{n,k_x,k_y,\sigma},\nonumber
\end{eqnarray}
so that
\begin{eqnarray}
\sum_{x,\sigma}\psi^*_{n',k_x',k_y,\sigma}(x)x^2\psi_{n,k_x,k_y,\sigma}(x)&=&
\sum_{x,\sigma}e^{i(k_x-k_x')x}u^*_{n',k_x',k_y,\sigma}\left[2ix
\frac{\partial}{\partial k_x}+\frac{\partial^2}{\partial k_x^2}\right]
u_{n,k_x,k_y,\sigma}\nonumber \\
&-&\frac{\partial^2}{\partial k_x^2}\sum_{x,\sigma}
\psi^*_{n',k_x',k_y,\sigma}(x)\psi_{n,k_x,k_y,\sigma}(x).
\end{eqnarray}
The first two terms are not manifestly Hermitian, however. If one had taken
derivatives with respect to $k_x'$ instead, one would have instead obtained
\begin{eqnarray}
\sum_{x,\sigma}\psi^*_{n',k_x',k_y,\sigma}(x)x^2\psi_{n,k_x,k_y,\sigma}(x)&=&
\sum_{x,\sigma}e^{i(k_x-k_x')x}
u_{n,k_x,k_y,\sigma}\left[-2ix\frac{\partial}{\partial k_x'}
+\frac{\partial^2}{\partial{k_x'}^2}\right]u^*_{n',k_x',k_y,\sigma}
\nonumber \\
&-&\frac{\partial^2}{\partial{k_x'}^2}\sum_{x,\sigma}
\psi^*_{n',k_x',k_y,\sigma}(x)\psi_{n,k_x,k_y,\sigma}(x).
\end{eqnarray}
The last sum gives $\delta(k_x-k_x')\delta_{n,n'}$ from Eq.~(\ref{eq:ortho}).
For the second term,
because the sum over $x$ is over all space, we can set $x=x'+R_x$, where
$R_x$ is a translation by an arbitrary number of unit cell lengths and $x'$ is
restricted to a single unit cell. Then
\begin{eqnarray}
\sum_{x,\sigma}e^{i(k_x-k_x')x}u^*_{n',k_x',k_y,\sigma}
\frac{\partial^2}{\partial k_x^2}u_{n,k_x,k_y,\sigma}
&=&\sum_{R_x}e^{i(k_x-k_x')R_x}\sum_{x',\sigma}e^{i(k_x-k_x')x'}
u^*_{n',k_x',k_y,\sigma}\frac{\partial^2}{\partial k_x^2}u_{n,k_x,k_y,\sigma}
\nonumber \\
&=&\delta(k_x-k_x')\sum_{x\in{\rm u.c.},\sigma}e^{i(k_x-k_x')x}
u^*_{n',k_x',k_y,\sigma}\frac{\partial^2}{\partial k_x^2}u_{n,k_x,k_y,\sigma}
\nonumber \\
&=&\delta(k_x-k_x')\sum_{x\in{\rm u.c.},\sigma}u^*_{n',k_x',k_y,\sigma}
\frac{\partial^2}{\partial k_x^2}u_{n,k_x,k_y,\sigma},
\end{eqnarray}
where the sum is now only over a single unit cell rather than over all space.
In the last line we have made use of the fact that $k_x=k_x'$. Inserting these
results into the previous expression gives
\begin{eqnarray}
\sum_{x,\sigma}\psi^*_{n',k_x',k_y,\sigma}(x)x^2\psi_{n,k_x,k_y,\sigma}(x)
&=&\delta(k_x-k_x')\frac{1}{2}\sum_{x\in{\rm u.c.,\sigma}}\left(
u^*_{n',k_x',k_y,\sigma}\frac{\partial^2}{\partial k_x^2}u_{n,k_x,k_y,\sigma}
+u_{n,k_x,k_y,\sigma}\frac{\partial^2}{\partial{k_x'}^2}
u^*_{n',k_x',k_y,\sigma}\right)\nonumber \\
&-&\delta_{n,n'}\frac{1}{2}\left(\frac{\partial^2}{\partial k_x^2}
+\frac{\partial^2}{\partial{k_x'}^2}\right)\delta(k_x-k_x')\nonumber \\
&+&\sum_{x,\sigma}e^{i(k_x-k_x')x}ix\left(u^*_{n',k_x',k_y,\sigma}
\frac{\partial}{\partial k_x}u_{n,k_x,k_y,\sigma}-u_{n,k_x,k_y,\sigma}
\frac{\partial}{\partial k_x'}u^*_{n',k_x',k_y,\sigma}\right)
\nonumber \\
&=&\delta(k_x-k_x')\frac{1}{2}\sum_{x\in{\rm u.c.,\sigma}}\left(
u^*_{n',k_x',k_y,\sigma}\frac{\partial^2}{\partial k_x^2}u_{n,k_x,k_y,\sigma}
+u_{n,k_x,k_y,\sigma}\frac{\partial^2}{\partial{k_x'}^2}
u^*_{n',k_x',k_y,\sigma}\right)\nonumber \\
&-&\delta_{n,n'}\frac{1}{2}\left(\frac{\partial^2}{\partial k_x^2}
+\frac{\partial^2}{\partial{k_x'}^2}\right)\delta(k_x-k_x')\nonumber \\
&+&\sum_{x,\sigma}e^{-ik_x'x}u^*_{n',k_x',k_y,\sigma}
\left(\frac{\partial}{\partial k_x}e^{ik_xx}\right)
\frac{\partial}{\partial k_x}u_{n,k_x,k_y,\sigma}\nonumber \\
&+&\sum_{x,\sigma}\left(\frac{\partial}{\partial k_x'}e^{-ik_x'x}\right)
e^{ik_xx}u_{n,k_x,k_y,\sigma}\frac{\partial}{\partial k_x'}
u^*_{n',k_x',k_y,\sigma},
\end{eqnarray}
which is manifestly Hermitian. Inserting this into the full expression for
$\langle x^2\rangle$ gives
\begin{eqnarray}
\left\langle x^2\right\rangle&=&
\sum_{n,n'}\sum_{k_x,k_x'}\sum_{k_y}a^*_{n'}(k_x',k_y,t)
a_n(k_x,k_y,t)
\delta(k_x-k_x')\sum_{\sigma}\frac{1}{2}\left(
u^*_{n',k_x',k_y,\sigma}\frac{\partial^2}{\partial k_x^2}u_{n,k_x,k_y,\sigma}
+u_{n,k_x,k_y,\sigma}\frac{\partial^2}{\partial k_x^2}u^*_{n',k_x',k_y,\sigma}
\right)\nonumber \\
&-&\frac{1}{2}\sum_{n,n'}\sum_{k_x,k_x'}\sum_{k_y}a^*_{n'}(k_x',k_y,t)
a_n(k_x,k_y,t)
\left(\frac{\partial^2}{\partial k_x^2}+\frac{\partial^2}{\partial{k_x'}^2}
\right)\delta(k_x-k_x')\delta_{n,n'}\nonumber \\
&+&\sum_{n,n'}\sum_{k_x,k_x'}\sum_{k_y}a^*_{n'}(k_x',k_y,t)
a_n(k_x,k_y,t)\sum_{x,\sigma}e^{-ik_x'x}
u^*_{n',k_x',k_y,\sigma}\left(\frac{\partial}{\partial k_x}e^{ik_xx}\right)
\frac{\partial}{\partial k_x}u_{n,k_x,k_y,\sigma}\nonumber \\
&+&\sum_{n,n'}\sum_{k_x,k_x'}\sum_{k_y}a^*_{n'}(k_x',k_y,t)
a_n(k_x,k_y,t)\sum_{x,\sigma}\left(\frac{\partial}{\partial k_x'}e^{-ik_x'x}
\right)u_{n,k_x,k_y,\sigma}\frac{\partial}{\partial k_x'}
u^*_{n',k_x',k_y,\sigma}e^{ik_xx}.
\end{eqnarray}
To evaluate the second line, we can use the identity
\begin{equation}\sum_xf(x)\delta^{(n)}(x-x_0)=(-1)^nf^{(n)}(x_0),
\end{equation}
which can be verified by integrating by parts $n$ times. Likewise, one can
integrate the last two terms by parts once. One then obtains
\begin{eqnarray}
\left\langle x^2\right\rangle
&=&\frac{1}{2}\sum_{n,n'}\sum_{\bm k}\left(a^*_{n'}({\bm k},t)a_n({\bm k},t)
\sum_{\sigma}u^*_{n',{\bm k},\sigma}\frac{\partial^2}{\partial k_x^2}
u_{n,{\bm k},\sigma}+{\rm H.c.}'\right)-\frac{1}{2}\sum_{n}\sum_{\bm k}\left(
a^*_{n}({\bm k},t)\frac{\partial^2}{\partial k_x^2}a_n({\bm k},t)+{\rm H.c.}'
\right)\nonumber \\
&-&\sum_{n,n'}\sum_{k_x,k_x'}\sum_{k_y}a^*_{n'}(k_x',k_y,t)\sum_{x,\sigma}
e^{i(k_x-k_x')x}u^*_{n',k_x',k_y,\sigma}\frac{\partial}{\partial k_x}\left(
a_n(k_x,k_y,t)\frac{\partial}{\partial k_x}u_{n,k_x,k_y,\sigma}\right)
\nonumber \\
&-&\sum_{n,n'}\sum_{k_x,k_x'}\sum_{k_y}a_n(k_x,k_y,t)\sum_{x,\sigma}
e^{i(k_x-k_x')x}u_{n,k_x,k_y,\sigma}\frac{\partial}{\partial k_x'}\left(
a^*_{n'}(k_x',k_y,t)\frac{\partial}{\partial k_x'}u^*_{n',k_x',k_y,\sigma}
\right)
\nonumber \\
&=&\frac{1}{2}\sum_{n,n'}\sum_{\bm k}\left(a^*_{n'}({\bm k},t)a_n({\bm k},t)
\sum_{\sigma}u^*_{n',{\bm k},\sigma}\frac{\partial^2}{\partial k_x^2}
u_{n,{\bm k},\sigma}+{\rm H.c.}'\right)-\frac{1}{2}\sum_{n}\sum_{\bm k}\left(
a^*_{n}({\bm k},t)\frac{\partial^2}{\partial k_x^2}a_n({\bm k},t)+{\rm H.c.}'
\right)\nonumber \\
&-&\sum_{n,n'}\sum_{{\bm k}}a^*_{n'}({\bm k},t)\sum_{\sigma}
u^*_{n',{\bm k},\sigma}\left(\frac{\partial}{\partial k_x}a_n({\bm k},t)
\frac{\partial}{\partial k_x}u_{n,{\bm k},\sigma}+a_n({\bm k},t)
\frac{\partial^2}{\partial k_x^2}u_{n,{\bm k},\sigma}\right)\nonumber \\
&-&\sum_{n,n'}\sum_{{\bm k}}a_n({\bm k},t)\sum_{\sigma}u_{n,{\bm k},\sigma}
\left(\frac{\partial}{\partial k_x}a^*_{n'}({\bm k},t)
\frac{\partial}{\partial k_x}u^*_{n',{\bm k},\sigma}+a^*_{n'}({\bm k},t)
\frac{\partial^2}{\partial k_x^2}u^*_{n',{\bm k},\sigma}\right)\nonumber \\
&=&-\frac{1}{2}\sum_{n,n'}\sum_{\bm k}\left(a^*_{n'}({\bm k},t)a_n({\bm k},t)
\sum_{\sigma}u^*_{n',{\bm k},\sigma}\frac{\partial^2}{\partial k_x^2}
u_{n,{\bm k},\sigma}+{\rm H.c.}'\right)-\frac{1}{2}\sum_{n}\sum_{\bm k}\left(
a^*_{n}({\bm k},t)\frac{\partial^2}{\partial k_x^2}a_n({\bm k},t)+{\rm H.c.}'
\right)\nonumber \\
&-&\sum_{n,n'}\sum_{{\bm k}}\left(a^*_{n'}({\bm k},t)
\frac{\partial}{\partial k_x}a_n({\bm k},t)\sum_{\sigma}
u^*_{n',{\bm k},\sigma}\frac{\partial}{\partial k_x}u_{n,{\bm k},\sigma}
+{\rm H.c.}'\right),
\end{eqnarray}
where the notation ${\rm H.c.}'$ represents complex conjugation of all
quantities as well as the interchange of band labels $n\leftrightarrow n'$
(these are dummy indices). The expectation of $r^2$ is therefore
\begin{eqnarray}
\langle r^2\rangle
&=&-\frac{1}{2}\sum_{n,n'}\sum_{\bm k}\left(a^*_{n'}({\bm k},t)a_n({\bm k},t)
\sum_{\sigma}u^*_{n',{\bm k},\sigma}\nabla_{\bm k}^2u_{n,{\bm k},\sigma}
+{\rm H.c.}'\right)-\frac{1}{2}\sum_{n}\sum_{\bm k}\left(a^*_{n}({\bm k},t)
\nabla_{\bm k}^2a_n({\bm k},t)+{\rm H.c.}'\right)\nonumber \\
&-&\sum_{n,n'}\sum_{{\bm k}}\left(a^*_{n'}({\bm k},t)
\nabla_{\bm k}a_n({\bm k},t)\cdot\sum_{\sigma}
u^*_{n',{\bm k},\sigma}\nabla_{\bm k}u_{n,{\bm k},\sigma}+{\rm H.c.}'\right).
\label{eq:r2exp3}
\end{eqnarray}
Recall that the Berry connections are defined as
\begin{equation}
A_i({\bm k})_{n'n}=\left\langle u_{n',{\bm k}}\bigg|i
\frac{\partial}{\partial k_i}\bigg|u_{n,{\bm k}}\right\rangle
=\sum_{\sigma}u^*_{n',{\bm k},\sigma}i\frac{\partial}{\partial k_i}
u_{n,{\bm k},\sigma}.
\label{eq:Berrypotentialx}
\end{equation}
We can therefore write Eq.~(\ref{eq:r2exp3}) explicitly in terms of the Berry
connections as follows:
\begin{eqnarray}
\langle r^2\rangle
&=&-\frac{1}{2}\sum_{n}\sum_{\bm k}\left(a^*_{n}({\bm k},t)
\nabla_{\bm k}^2a_n({\bm k},t)+{\rm H.c.}'\right)
-\frac{1}{2}\sum_{n,n'}\sum_{\bm k}\left(a^*_{n'}({\bm k},t)a_n({\bm k},t)
\sum_{\sigma}u^*_{n',{\bm k},\sigma}\nabla_{\bm k}^2u_{n,{\bm k},\sigma}
+{\rm H.c.}'\right)
\nonumber \\
&+&i\sum_{n,n'}\sum_{{\bm k}}\left(a^*_{n'}({\bm k},t)
\nabla_{\bm k}a_n({\bm k},t)\cdot{\bm A}({\bm k})_{n'n}-{\rm H.c.}'\right).
\label{eq:r2exp4}
\end{eqnarray}

\mm The time-evolution of the state~(\ref{eq:psidef1}) is given by
the solution of the time-dependent Schr\" odinger equation,
\begin{equation}
|\psi({\bm r},t)\rangle=\sum_n\sum_{{\bm k}}a_n({\bm k},t)
e^{i{\bm k}\cdot{\bm r}}|u_{n,{\bm k}}\rangle=\sum_n\sum_{{\bm k}}
e^{i{\bm k}\cdot{\bm r}}a_n({\bm k},0)e^{-iE_n({\bm k})t/\hbar}
|u_{n,{\bm k}}\rangle
\end{equation}
or
\begin{equation}
\psi_{\sigma}({\bm r},t)\rangle=\sum_n\sum_{{\bm k}}e^{i{\bm k}\cdot{\bm r}}
a_n({\bm k},0)e^{-iE_n({\bm k})t/\hbar}u_{n,{\bm k},\sigma},
\label{eq:psidef3}
\end{equation}
so that one may consider the time evolution to be driven entirely by the
amplitudes:
\begin{equation}
a_n({\bm k},t)=e^{-iE_n({\bm k})t/\hbar}a_n({\bm k},0)\equiv
e^{-iE_n({\bm k})t/\hbar}a_n({\bm k}).
\end{equation}
Then $a^*_{n'}({\bm k},t)a_n({\bm k},t)
=e^{-i\left[E_n({\bm k})-E_{n'}({\bm k})\right]t/\hbar}a^*_{n'}({\bm k})
a_n({\bm k})$,
\begin{equation}
a^*_{n'}({\bm k},t)\frac{\partial}{\partial k_i}a_n({\bm k},t)
=a^*_{n'}({\bm k})e^{-i\left[E_n({\bm k})-E_{n'}({\bm k})\right]t/\hbar}\left(
-\frac{it}{\hbar}\frac{\partial E_n({\bm k})}{\partial k_i}a_n({\bm k})+
\frac{\partial a_n({\bm k})}{\partial k_i}\right)
\end{equation}
and
\begin{eqnarray}
a^*_{n'}({\bm k},t)\frac{\partial^2}{\partial k_i^2}a_n({\bm k},t)
&=&a^*_{n'}({\bm k})e^{iE_{n'}({\bm k})t/\hbar}\bigg[
-\frac{it}{\hbar}\frac{\partial^2E_n({\bm k})}{\partial k_i^2}a_n({\bm k})
-\frac{it}{\hbar}\frac{\partial E_n({\bm k})}{\partial k_i}
\frac{\partial}{\partial k_i}a_n({\bm k})
-\frac{t^2}{\hbar^2}\left(\frac{\partial E_n({\bm k})}{\partial k_i}\right)^2
a_n({\bm k})\nonumber \\
&+&\frac{\partial^2}{\partial k_i^2}a_n({\bm k})
-\frac{it}{\hbar}\frac{\partial E_n({\bm k})}{\partial k_i}
\frac{\partial}{\partial k_i}a_n({\bm k})\bigg]
e^{-iE_n({\bm k})t/\hbar}\nonumber \\
&=&a^*_{n'}({\bm k})e^{-i\left[E_n({\bm k})-E_{n'}({\bm k})\right]t/\hbar}
\bigg[
-\frac{it}{\hbar}\frac{\partial^2 E_n({\bm k})}{\partial k_i^2}a_n({\bm k})
-\frac{2it}{\hbar}\frac{\partial E_n({\bm k})}{\partial k_i}
\frac{\partial a_n({\bm k})}{\partial k_i}\nonumber \\
&&\qquad -\frac{t^2}{\hbar^2}\left(\frac{\partial E_n({\bm k})}{\partial k_i}
\right)^2a_n({\bm k})+\frac{\partial^2a_n({\bm k})}{\partial k_i^2}\bigg].
\end{eqnarray}
Inserting these into Eq.~(\ref{eq:r2exp4}) gives
\begin{eqnarray}
\langle r^2\rangle
&=&
-\frac{1}{2}\sum_{n}\sum_{\bm k}\Bigg[
a^*_n({\bm k})\bigg(-\frac{it}{\hbar}\nabla_{\bm k}^2E_n({\bm k})a_n({\bm k})
-\frac{2it}{\hbar}\nabla_{\bm k}E_n({\bm k})\cdot\nabla_{\bm k}a_n({\bm k})
\nonumber \\
&&\qquad -\frac{t^2}{\hbar^2}\left(\nabla_{\bm k}E_n({\bm k})\right)^2
a_n({\bm k})+\nabla_{\bm k}^2a_n({\bm k})\bigg)+{\rm H.c.}'\Bigg]\nonumber \\
&-&\frac{1}{2}\sum_{n,n'}\sum_{\bm k}\left(
e^{-i\left[E_n({\bm k})-E_{n'}({\bm k})\right]t/\hbar}a^*_{n'}({\bm k})
a_n({\bm k})\sum_{\sigma}u^*_{n',{\bm k},\sigma}\nabla_{\bm k}^2
u_{n,{\bm k},\sigma}+{\rm H.c.}'\right)\nonumber \\
&-&\sum_{n,n'}\sum_{{\bm k}}\left(a^*_{n'}({\bm k})
e^{-i\left[E_n({\bm k})-E_{n'}({\bm k})\right]t/\hbar}\left(
-\frac{it}{\hbar}a_n({\bm k})\nabla_{\bm k}E_n({\bm k})
+\nabla_{\bm k}a_n({\bm k})\right)
\cdot{\bm B}_{n'n}+{\rm H.c.}'\right),
\label{eq:r2exp5}
\end{eqnarray}
where ${\bm B}_{n'n}\equiv{\bm A}_{n'n}/i$ is defined to avoid mistakes when 
Hermitian conjugating various terms. The first term in square brackets cancels 
with its Hermitian conjugate, while the third term is equal to its conjugate. 
At long times, this is the only term that is relevant as it has the largest 
($t^2$) time-dependent prefactor. We are interested in the time-derivative of 
the (time-dependent) variance (note that $\langle r\rangle=0$):
\begin{equation}
\frac{\partial}{\partial t}\langle r^2\rangle.
\label{eq:dvardt}
\end{equation}
The only terms in Eq.~(\ref{eq:r2exp5}) that contribute to
Eq~(\ref{eq:dvardt}) are those that explicitly depend on time, so one may write
\begin{eqnarray}
\frac{\partial}{\partial t}
\langle r^2\rangle
&=&\frac{t}{\hbar^2}\sum_{n}\sum_{\bm k}|a_n({\bm k})|^2
\left(\nabla_{\bm k}E_n({\bm k})\right)^2
+\sum_{n}\sum_{\bm k}\left(\frac{i}{\hbar}a^*_n({\bm k})\nabla_{\bm k}
E_n({\bm k})\cdot\nabla_{\bm k}a_n({\bm k})+{\rm H.c.}'\right)\nonumber \\
&+&\frac{1}{2}\sum_{n,n'}\sum_{\bm k}\left(\frac{i}{\hbar}\left(E_n({\bm k})
-E_{n'}({\bm k})\right)e^{-i\left[E_n({\bm k})-E_{n'}({\bm k})\right]t/\hbar}
a^*_{n'}({\bm k})a_n({\bm k})\sum_{\sigma}u^*_{n',{\bm k},\sigma}
\nabla_{\bm k}^2u_{n,{\bm k},\sigma}+{\rm H.c.}'\right)\nonumber \\
&+&\sum_{n,n'}\sum_{{\bm k}}\frac{i}{\hbar}\left(E_n({\bm k})-E_{n'}({\bm k})
\right)a^*_{n'}({\bm k})
e^{-i\left[E_n({\bm k})-E_{n'}({\bm k})\right]t/\hbar}\left(
-\frac{it}{\hbar}a_n({\bm k})\nabla_{\bm k}E_n({\bm k})
+\nabla_{\bm k}a_n({\bm k})\right)\cdot{\bm B}_{n'n}\nonumber \\
&+&\sum_{n,n'}\sum_{{\bm k}}\frac{i}{\hbar}a^*_{n'}({\bm k})a_n({\bm k})
e^{-i\left[E_n({\bm k})-E_{n'}({\bm k})\right]t/\hbar}\nabla_{\bm k}E_n({\bm k})
\cdot{\bm B}_{n'n}\nonumber \\
&-&\sum_{n,n'}\sum_{{\bm k}}\frac{i}{\hbar}\left(
E_n({\bm k})-E_{n'}({\bm k})\right)a_{n'}({\bm k})
e^{i\left[E_n({\bm k})-E_{n'}({\bm k})\right]t/\hbar}\left(
\frac{it}{\hbar}a^*_n({\bm k})\nabla_{\bm k}E_n({\bm k})
+\nabla_{\bm k}a^*_n({\bm k})\right)\cdot{\bm B}^*_{n'n}\nonumber \\
&-&\sum_{n,n'}\sum_{{\bm k}}\frac{i}{\hbar}a_{n'}({\bm k})a^*_n({\bm k})
e^{i\left[E_n({\bm k})-E_{n'}({\bm k})\right]t/\hbar}\nabla_{\bm k}E_n({\bm k})
\cdot{\bm B}^*_{n'n}.
\label{eq:dr2dt}
\end{eqnarray}
At long times, only the terms proportional to $t$ are relevant, so one may 
write
\begin{eqnarray}
\frac{\partial}{\partial t}
\langle r^2\rangle
&\approx&\frac{t}{\hbar^2}\sum_{n}\sum_{\bm k}|a_n({\bm k})|^2
\left(\nabla_{\bm k}E_n({\bm k})\right)^2\nonumber \\
&+&\frac{t}{\hbar^2}\sum_{n,n'}\sum_{{\bm k}}\left(E_n({\bm k})-E_{n'}({\bm k})
\right)a^*_{n'}({\bm k})a_n({\bm k})
e^{-i\left[E_n({\bm k})-E_{n'}({\bm k})\right]t/\hbar}
\nabla_{\bm k}E_n({\bm k})\cdot{\bm B}_{n'n}\nonumber \\
&+&\frac{t}{\hbar^2}\sum_{n,n'}\sum_{{\bm k}}\left(E_n({\bm k})-E_{n'}({\bm k})
\right)a_{n'}({\bm k})a^*_n({\bm k})
e^{i\left[E_n({\bm k})-E_{n'}({\bm k})\right]t/\hbar}
\nabla_{\bm k}E_n({\bm k})\cdot{\bm B}^*_{n'n}.
\label{eq:dr2dt2}
\end{eqnarray}
It is useful to expand this explicitly in band index, keeping in mind that
$E_+({\bm k})\equiv E_{\bm k}$ and $E_-({\bm k})\equiv -E_{\bm k}$:
\begin{eqnarray}
\frac{\partial}{\partial t}
\langle r^2\rangle
&\approx&\frac{t}{\hbar^2}\sum_{\bm k}\left(\nabla_{\bm k}E_{\bm k}\right)^2
\nonumber \\
&+&\frac{2t}{\hbar^2}\sum_{{\bm k}}E_{\bm k}a^*_+({\bm k})a_-({\bm k})
e^{2iE_{\bm k}t/\hbar}\nabla_{\bm k}E_{\bm k}\cdot{\bm B}_{+-}
+\frac{2t}{\hbar^2}\sum_{{\bm k}}E_{\bm k}a_+({\bm k})a^*_-({\bm k})
e^{-2iE_{\bm k}t/\hbar}\nabla_{\bm k}E_{\bm k}\cdot{\bm B}^*_{+-}\nonumber \\
&+&\frac{2t}{\hbar^2}\sum_{{\bm k}}E_{\bm k}a^*_-({\bm k})a_+({\bm k})
e^{-2iE_{\bm k}t/\hbar}\nabla_{\bm k}E_{\bm k}\cdot{\bm B}_{-+}
+\frac{2t}{\hbar^2}\sum_{{\bm k}}E_{\bm k}a_-({\bm k})a^*_+({\bm k})
e^{2iE_{\bm k}t/\hbar}\nabla_{\bm k}E_{\bm k}\cdot{\bm B}^*_{-+}\nonumber \\
&=&\frac{t}{\hbar^2}\sum_{\bm k}\left(\nabla_{\bm k}E_{\bm k}\right)^2
+\frac{2t}{\hbar^2}\sum_{{\bm k}}E_{\bm k}a^*_+({\bm k})a_-({\bm k})
e^{2iE_{\bm k}t/\hbar}\nabla_{\bm k}E_{\bm k}\cdot\left({\bm B}_{+-}+{\bm B}^*_{-+}\right)
\nonumber \\
&+&\frac{2t}{\hbar^2}\sum_{{\bm k}}E_{\bm k}a_+({\bm k})a^*_-({\bm k})
e^{-2iE_{\bm k}t/\hbar}\nabla_{\bm k}E_{\bm k}\cdot\left({\bm B}^*_{+-}+{\bm B}_{-+}\right).
\label{eq:dr2dt2b}
\end{eqnarray}
We know that $(A_i)_{n'n}=(A_i)^*_{nn'}$, but also that $A_i=iB_i$, which gives
$(iB_i)_{n'n}=i(B_i)_{n'n}=(iB_i)^*_{nn'}=-i(B_i)^*_{nn'}$ or
$(B_i)_{n'n}=-(B_i)^*_{nn'}$. This means that both
${\bm B}_{+-}+{\bm B}^*_{-+}=0$ and
${\bm B}_{-+}+{\bm B}^*_{+-}=0$, i.e.\ that the last two terms in the above
equation vanish identically. Therefore, at long times the time-derivative of
the variance depends only on the Hamiltonian, and not on the Berry
connections:
\begin{equation}
\frac{\partial}{\partial t}\langle r^2\rangle
\approx\frac{t}{\hbar^2}\sum_{\bm k}\left(\nabla_{\bm k}E_{\bm k}\right)^2.
\end{equation}

\end{widetext}

\end{document}